\documentclass[aps,pra,superscriptaddress]{revtex4}
\usepackage{longtable}
\usepackage{graphicx}
\usepackage{times}
\usepackage{amssymb}
\newcommand{\beq}{\begin{equation}}
\newcommand{\eeq}{\end{equation}}
\newcommand{\bea}{\begin{eqnarray}}
\newcommand{\eea}{\end{eqnarray}}
\newcommand{\vectwo}[2]{\left(\begin{array}{c}#1 \\ #2 \end{array} \right)}

\begin{document}


\title{Squeezed-input, optical-spring, signal-recycled gravitational-wave detectors}

\author{Jan Harms}
\affiliation{Max-Planck-Institut f\"ur Gravitationsphysik, Albert-Einstein-Institut, Universit\"at Hannover, Callinstr. 38, 30167 Hannover, Germany}

\author{Yanbei Chen}
\affiliation{Theoretical Astrophysics 130-33, California Institute of Technology, Pasadena, California 91125, USA}

\author{Simon Chelkowski}
\affiliation{Max-Planck-Institut f\"ur Gravitationsphysik, Albert-Einstein-Institut, Universit\"at Hannover, Callinstr. 38, 30167 Hannover, Germany}

\author{Alexander Franzen}
\affiliation{Max-Planck-Institut f\"ur Gravitationsphysik, Albert-Einstein-Institut, Universit\"at Hannover, Callinstr. 38, 30167 Hannover, Germany}

\author{Henning Vahlbruch}
\affiliation{Max-Planck-Institut f\"ur Gravitationsphysik, Albert-Einstein-Institut, Universit\"at Hannover, Callinstr. 38, 30167 Hannover, Germany}

\author{Karsten Danzmann}
\affiliation{Max-Planck-Institut f\"ur Gravitationsphysik, Albert-Einstein-Institut, Universit\"at Hannover, Callinstr. 38, 30167 Hannover, Germany}

\author{Roman Schnabel}
\affiliation{Max-Planck-Institut f\"ur Gravitationsphysik, Albert-Einstein-Institut, Universit\"at Hannover, Callinstr. 38, 30167 Hannover, Germany}




\date{\today}

\begin{abstract}
We theoretically analyze the quantum noise of signal-recycled laser interferometric gravitational-wave detectors with additional input and output optics, namely frequency-dependent squeezing of the vacuum state of light entering the dark port and frequency-dependent homodyne detection. We combine the work of Buonanno and Chen on the quantum noise of signal-recycled interferometers with ordinary input and output optics, and the work of Kimble {\it el al.} on frequency-dependent input and output optics with conventional interferometers. Analytical formulas for the optimal input and output frequency dependencies are obtained. It is shown that injecting squeezed light with the optimal frequency-dependent squeezing angle into the dark port yields an improvement on the noise spectral density by a factor of $e^{-2r}$ (in power) over the entire squeezing bandwidth, where $r$ is the squeezing parameter. It is further shown that frequency-dependent (variational) homodyne read-out leads to an additional increase in sensitivity which is significant in the wings of the doubly resonant structure. The optimal variational input squeezing in case of an ordinary output homodyne detection is shown to be realizable by applying two optical filters on a  frequency-independent squeezed vacuum. Throughout this paper, we take as example the signal-recycled  topology currently being completed at the GEO\,600 site. However, theoretical results obtained here are also  applicable to the proposed topology of {\it Advanced LIGO}.
\end{abstract}

\maketitle

\section{introduction}

Gravitational waves (GW) have long been predicted by Albert Einstein using the theory of general relativity, but so far have not been directly observed \cite{Thorne87}. 
Currently, an international array of first-generation, kilometer-scale laser interferometric gravitational-wave detectors, consisting of GEO\,600~\cite{geo02}, LIGO~\cite{LIGO}, TAMA~300~\cite{TAMA} and VIRGO~\cite{VIRGO}, targeted at gravitational-waves in the acoustic band from 10~Hz  to 10~kHz, is going into operation.\
These first-generation detectors are all Michelson interferometers with suspended mirrors. Injecting a strong carrier light from the bright port, the anti-symmetric mode of arm-lengths oscillations (e.g. excited by a gravitational wave) yields a sideband modulation field in the anti-symmetric (optical) mode which is detected at the dark output port.
To yield a high sensitivity to gravitational waves, long arm lengths of 300~m up to 4~km and circulating laser power in the order of 10~kW are going to be realized in 2003 with the help of the technique of {\it power recycling}~\cite{DHKHFMW83pr}. 

GEO\,600 is the only first-generation detector that not only uses power recycling, but also includes the more advanced technique of {\it signal recycling}~\cite{Mee88}. The idea of signal recycling is to retro-reflect part of the signal light at the dark port back into the interferometer, establishing an additional cavity which can be set to resonate at a desired gravitational-wave frequency. Signal recycling leads to a well known (optical) resonance structure in the interferometer's sensitivity curve. This resonance can already beat the standard quantum limit (SQL) \cite{BCh01a,BCh01b}, which is the upper bound for the sensitivity of conventional interferometers without signal recycling and with conventional input and output optics. A further benefit of signal recycling is the reduced optical loss due to imperfect mode matching from the {\em mode healing effect}~\cite{SMe91}. The next-generation detectors currently being planned are  likely to use this technique, for example the {\it Advanced} LIGO (LIGO II)~\cite{GSSW99}.

Buonanno and Chen also predict a second, opto-mechanical resonance in signal-recycled interferometers, around which the interferometer gains sensitivity, and can also beat the standard quantum limit~\cite{BCh01a,BCh01b,BCh02a,BCh03a}. Their work has been limited to signal-recycled interferometers with arm cavities, or interferometers with one single end mirror in each arm, and with infinitely heavy beamsplitters. In all cases considered, coherent vacuum was entering the interferometer's dark port, i.e. no additional input and output optics were investigated.
On the other hand, Kimble {\it et al.} investigated these additional input and output optics for the conventional LIGO detector topology without signal-recycling \cite{KLMTV01} building on earlier work on squeezed-input interferometers \cite{Cav81,Unruh82,GLe87,JRe90,PCW93} and variational-output interferometers \cite{VMa93,VZu95,VMa96a,VMa96b,VZu98}. 

In this paper, we investigate the benefit of squeezed light with frequency-dependent squeezing angle injected into the interferometer's dark port and also the benefit of frequency-dependent (variational) homodyne readout, using the two-photon input-output formalism of quantum optics \cite{CSc85} . 
In sections \ref{SI} and \ref{VO} of this paper we derive analytical expressions for the optimized frequency dependencies of squeezing angle and homodyning angle for optical-spring signal-recycled interferometers, respectively. For definiteness, our results are presented using the Michelson topology of GEO\,600. Unlike the LIGO, VIRGO and TAMA\,300 interferometers, GEO\,600 has folded arms and no arm cavities (Fig. \ref{GEO600}). We plot and compare the spectral densities of the quantum noise of the GEO\,600 topology without and with additional input and output optics. Using the coupling parameter of Advanced LIGO, the results are readily applicable to the proposed LIGO topology.

\section{Signal recycling}
\label{SR}

\begin{figure}[ht!]
  \centerline{\includegraphics[angle=-90,width=9.0cm]{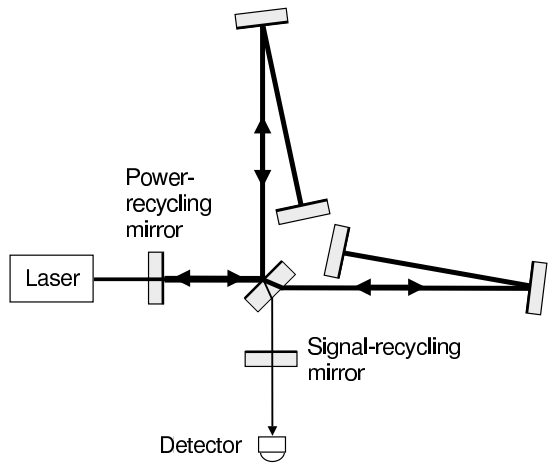}}
  \vspace{0mm}
  \caption{GEO\,600 is a dual-recycled Michelson interferometer implementing a power-recycling mirror in order to enhance the light power within the Michelson arms and a signal-recycling mirror in the dark port which is tuned on a specific signal frequency. Since the arms are folded once, then the effective armlength is doubled to 1200~m.}
  \label{GEO600}
\end{figure}

By placing a mirror in the dark port of an interferometer a cavity between this so-called signal-recycling mirror and the two end mirrors of the interferometer is formed. The length of this cavity can be tuned independently and can be made resonant at some signal frequency $\Omega$. Thus the signal is recycled and amplified due to an increased interaction time. 
The original idea of the signal-recycling (SR) topology, i.e. a mirror in the dark port, was due to Meers \cite{Mee88}, who proposed its use for {\it dual-recycling}, which is the combination of power- and signal-recycling. Later, Mizuno {\it et al.} \cite{MSNCSRWD93,Mizuno95} 
and Heinzel {\it et al.} \cite{Heinzel99,HMSRWD96} proposed the scheme of {\it resonant sideband extraction}, which uses a detuned signal-recycling mirror to extract the signal from high-finesse arm cavities. Both schemes of tuned and detuned signal-recycling cavities have been experimentally demonstrated by Heinzel {\it et al.} \cite{HSMSWWSRD98} and Freise {\it et al.} \cite{FHSMSLWSRWD00} with the 30~m laser interferometer in Garching near Munich. Recently the GEO\,600 interferometer in Ruthe near Hannover has been completed by the implementation of the signal-recycling mirror. Since GEO\,600 has no Fabry-P\'erot cavities (Fig. \ref{GEO600})  the SR-mirror will be operated at or close to resonance. Relevant technical parameters of GEO\,600 are summerized in Table \ref{GEOdata}.

\begin{table}[ht!]
\begin{center}
\begin{tabular}{|c|l|l|}
\hline
~symbol~& ~physical meaning	& ~numerical value \rule{0mm}{5mm}\\[0.5mm]
\hline
$m$ 	& ~mirror mass (each)	&~ $5.6\,{\rm kg}$	\\
$L$ 	& ~effective arm length 	&~ $1200\,{\rm m}$	\\
$P$ 	& ~circulating light power &~ $10\,{\rm kW}$ 	\\
$\omega_0$ & ~angular frequency of carrier light &~ $1.77\cdot10^{15}\,{\rm s}^{-1}$ \\
$\rho$ & ~power reflectivity of SRM &~ $0.99$	 \\
$\phi$& ~SR-cavity detuning 	&~ $0.0055\,\rm{rad}$ \\
\hline
\end{tabular}
\caption[Constants]{\it Technical data and parameter values of GEO\,600 which where used to calculate the spectral noise densities in Figs.~\ref{noise1}--\ref{noise3}.}
\label{GEOdata}
\end{center}
\end{table}

Whereas it was well known that signal-recycled interferometers exhibit an optical resonance, Buonanno and Chen \cite{BCh01a,BCh01b} have recently shown that SR-interferometers exhibit a second resonance, which is opto-mechanical. This resonance stems from the classical opto-mechanical coupling of the light field with the anti-symmetric mode of the otherwise free mirrors~\cite{BCh02a}: in detuned signal-recycling schemes, the phase-modulation sidebands induced by a gravitational wave are partly converted into amplitude modulations, which beat with the carrier field, producing a motion-dependent force and acting back on the test masses. This classical back-action force can be thought of as generated by an {\it optical spring}. The optical spring makes the test masses no longer free, and can shift their resonant frequencies upwards into the detection band. The interferometer gains sensitivity on and around this resonance, and can beat the standard quantum limit~\cite{BCh01a,BCh01b,BCh02a}. Whereas the optical resonance is primarily determined by the detuning of the SR-cavity with respect to the carrier-frequency $\omega_0$ (Fig.~5.8 in \cite{HarmsDipl}), the opto-mechanical resonance appears at a specific sideband frequency of the carrier light which depends on the interferometer's topology, the mirror masses $m$, the light-power $P$ inside the interferometer and the detuning $\phi$ of the SR cavity from its resonance. The opto-mechanical coupling of the light field with the anti-symmetric mode of the interferometer also leads to the phenomenon of ponderomotive squeezing \cite{BMa67}, i.e. the amplitude and phase quantum noise become correlated. This quantum effect is automatically considered by the formalism revealing the optical-spring behaviour. However, as pointed out in \cite{BCh01b}, in SR interferometers the ponderomotive squeezing only seems to be a secondary factor that enables the interferometer to beat the SQL, whereas the {\it classical} resonant amplification of the signal provides the main factor.

The investigations led by Buonanno and Chen focused on the topology of the proposed {\it Advanced} LIGO configuration, which consists of a dual-recycled Michelson-interferometer with a Fabry-P\'erot cavity in each arm. Due to the weak laser power at the beamsplitter, the opto-mechanical coupling of the light with the beamsplitters free oscillation was neglected.
In contrast, GEO\,600 is a dual-recycled interferometer that builds up a high intensity field by means of a power-recycling (PR) mirror in the bright port of the interferometer. Therefore, the motion of the beamsplitter (BS) in GEO\,600 is affected by power fluctuations of fields impinging from different directions. Nevertheless, assuming that the laser is shot-noise limited, the opto-mechanical coupling at the beamsplitter exerts only minor changes on the noise spectrum of the output. It can intuitively be understood that the quantum back-action noises associated with the arm mirrors, which have a reduced mass of $1/5$ the actual mirror mass due to folding the arms, clearly dominates the beamsplitter of $m_{\rm BS}=9.3\,{\rm kg}$.
Throughout this paper we do not consider the effect of radiation-pressure noise on the beamsplitter. This has also been studied but will be presented in detail elsewhere \cite{Harms03b}. Henceforth the term ``ideal GEO\,600'' refers to the interferometer with opto-mechanical coupling of the beamsplitter neglected.

The optical noise in an interferometer can be expressed in terms of the (single-sided) noise spectral density $S_h$ of the output field normalized by the transfer function of the signal. The noise spectral density is obtained from the {\it input-output relation}, which maps the numerous input fields $\overline{\mathbf{i}}_n$ and the gravitational-wave signal $h=\Delta L/L$ onto the detected output field $\overline{\mathbf{o}}$. Here we note that no additional noise due to the quantization of the test masses has to be considered. The sole forms of quantum noise affecting the output noise in interferometric gravitational wave detectors are the shot noise and the radiation pressure noise \cite{BGKMTV02}.

The following calculations are most easily accomplished in the Caves-Schumaker two-photon formalism~\cite{CSc85}, where the optical fields are decomposed into amplitude and phase quadratures, which can then be put together into a vector, e.g., for the output field of the interferometer
\beq
\overline\mathbf{o}=\vectwo{\hat{o}_1}{\hat{o}_2}\,,
\eeq
where $\hat{o}_{1,2}$ are the output amplitude and phase quadratures. The input-output relation for a lossless SR interferometer can be cast into the following form:
\beq
\overline{\mathbf{o}}=\frac{1}{M}\left[\mathbf{T}\overline{\mathbf{i}}
 + \overline{\mathbf{s}}\,h\right].
\label{IOrel}
\eeq
Here, $\mathbf{T}$ designates a $2\times 2$-matrix. Its four components are \cite{HarmsDipl}:
\beq
\begin{array}{rcl}
T_{\rm 11,22} & = & \rm{e}^{2\imath\Phi}\left[(1+\rho^2)\left(\cos(2\phi)+\frac{\mathcal{K}}{2}\sin(2\phi)\right)-2\rho\cos(2\Phi)\right],\\ \\
T_{12} & = & -\rm{e}^{2\imath\Phi}\tau^2\left(\sin(2\phi)+\mathcal{K}\sin^2(\phi)\right),\\ \\
T_{21} & = & \rm{e}^{2\imath\Phi}\tau^2\left(\sin(2\phi)-\mathcal{K}\cos^2(\phi)\right),
\end{array}
\label{TransM}
\eeq
and $M$ is given by
\beq
M=1+\rho^2\rm{e}^{4\imath\Phi}-2\rho\rm{e}^{2\imath\Phi}\left(\cos(2\phi)+\frac{\mathcal{K}}{2}\sin(2\phi)\right).
\eeq
Thus, $\mathbf{T}$ contains an overall phase factor $\rm{e}^{2\imath\Phi}$. $\rho$ and $\tau$ denote the amplitude reflectivity and transmissivity of the SR mirror. The signal transfer functions $\overline{\mathbf{s}}$ for the two quadratures are given by:
\beq
\begin{array}{rcl}
\hat{s}_1 & = & -\displaystyle{\frac{\sqrt{2\mathcal{K}}}{h_{\rm SQL}}}\,\tau\,\left(1+\rho\,\rm{e}^{2\imath\,\Phi}\right)\sin(\phi), \\ \\
\hat{s}_2 & = & -\displaystyle{\frac{\sqrt{2\mathcal{K}}}{h_{\rm SQL}}}\,\tau\,\left(-1+\rho\,\rm{e}^{2\imath\,\Phi}\right)\cos(\phi). 
\end{array}
\eeq
Remarkably, the input-output relations are formally identical for both configurations, {\it Advanced} LIGO and ideal GEO600. Their distinguishing properties lie in the definition of the opto-mechanical coupling-constant $\mathcal{K}$, the standard quantum-limit $h_{\rm SQL}$ and the phase-angle $\Phi$ which are also functions of the modulation-frequency $\Omega$.\\
\begin{table}
\begin{center}
\begin{tabular}{|c|c|c|}
\hline
{\,symbol\,}&{\,\,\,\,\,GEO\,600\,\,\,\,\,}&{\,Advanced LIGO}\rule{0mm}{5mm}\\[2mm]
\hline
$\mathcal{K}$ & $\displaystyle{\frac{20P\omega_0}{m\rm{c}^2\Omega^2}}$ & $\displaystyle{\frac{8P\omega_0}{mL^2\Omega^2(\Omega^2+\gamma_{\rm arm}^2)}}$ \rule{0mm}{7mm}\\[8mm]
\hline
$h_{\rm SQL}$ & $\displaystyle{\frac{20\hbar}{m\Omega^2L^2}}$ & $\displaystyle{\frac{8\hbar}{m\Omega^2L^2}}$ \rule{0mm}{5mm}\\[3mm]
\hline
$\Phi$ & $\displaystyle{\frac{\Omega L}{c}}$ & $\displaystyle{{\rm arctan}\left[\displaystyle{\frac{\Omega}{\gamma_{\rm arm}}}\right]}$ \rule{0mm}{6mm}\\[3mm]
\hline
\end{tabular}
\caption{Definitions of $\mathcal{K}$ \cite{Knote}, $h_{\rm SQL}$ and $\Phi$ for GEO\,600 and Advanced LIGO topologies. Here $m$ is the individual mirror mass, $L$ the Michelson arm length, $P$ the input power at the beamsplitter, $\omega_0$ the laser angular frequency, $\gamma_{\rm arm}=\tau_{\rm arm}^2c/(4L)$ the half linewidth of the Advanced LIGO arm cavity ($\tau_{\rm arm}$ the input test-mass mirror amplitude transmissivity), and $\Omega$ the GW sideband angular frequency. Values for Advanced LIGO are kept to the leading order of $\tau_{\rm arm}^2$, as in Refs.~\cite{BCh01a,BCh01b,BCh02a}.}
\end{center}
\label{SRparam}
\end{table}
A phase-sensitive measurement (i.e. homodyne or heterodyne) yields a photocurrent which depends linearly on a certain combination of the two output quadrature-fields:
\beq
\hat{o}_\zeta=\hat{o}_1\cos\zeta + \hat{o}_2\sin\zeta,
\eeq
where $\zeta$ is the homodyne angle (i.e. the angle of homodyne detection). The radiation-pressure forces acting on the mirrors are proportional to the amplitude quadrature and the motion-induced sideband fields are excitations of the light's phase quadrature.
\begin{figure}[ht!]
  \centerline{\includegraphics[width=8.0cm]{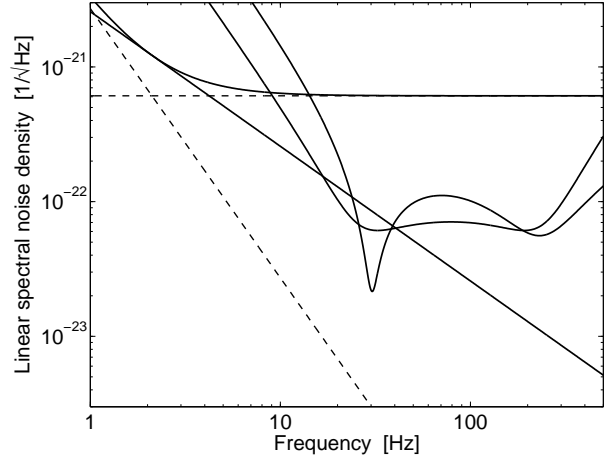}}
  \vspace{0mm}
  \caption{The dashed lines represent the uncorrelated white shot noise and the radiation-pressure noise ($\propto f^{-2}$). The sum of these shot and radiation-pressure noises yields the noise spectral density of a simple Michelson without arm cavities, here using GEO\,600 parameters and $\rho=0$.
In comparison, the noise spectral densities of both orthogonal quadratures of the signal-recycled GEO\,600 output-field exhibit a doubly resonant structure which beats the standard quantum limit.}
  \label{noise1}
\end{figure}
The noise spectral density when detecting the quadrature $\hat{o}_\zeta$ is determined by the transfer matrix $\mathbf{T}$ and the signal transfer-functions $\overline{\mathbf{s}}$. It assumes the form [see, e.g., Ref.~\cite{BCh01a}],
\beq
\label{S_h}
S_h=
\frac{\left(\begin{array}{cc}\cos\zeta & \sin\zeta\end{array}\right)
\mathbf{T}\mathbf{T}^{\dagger}\left(\begin{array}{c}\cos\zeta \\ \sin\zeta\end{array}\right)}
{\left(\begin{array}{cc}\cos\zeta & \sin\zeta\end{array}\right)
\overline{\mathbf{s}}\,\overline{\mathbf{s}}^{\dagger}
\left(\begin{array}{c}\cos\zeta \\ \sin\zeta\end{array}\right)},
\label{SRSD}
\eeq
provided that the input field $\overline{\mathbf{i}}$ entering from the dark port is a coherent vacuum field. Since $\overline{\mathbf{s}}$ is a complex vector, the product $\overline{\mathbf{s}}\,\overline{\mathbf{s}}^\dagger$ represents a symmetrized product 
\beq
\left<\overline{\mathbf{s}}\,\overline{\mathbf{s}}^\dagger\right>_{\rm{sym}}=\frac{1}{2}(\overline{\mathbf{s}}\,\overline{\mathbf{s}}^\dagger+\overline{\mathbf{s}}^\ast\overline{\mathbf{s}}^{\rm{T}}).
\eeq
The same holds for the matrix product $\mathbf{T}\mathbf{T}^{\dagger}$ in the nominator. Its symmetrization becomes necessary, if a more general interferometer topology is considered with complex coupling constant $\mathcal{K}$. The expression in Eq.~(\ref{SRSD}) for the noise spectral density is valid for any optical system whose transfer function can be given the form of Eq.~(\ref{IOrel}). 

Using Eq.~(\ref{S_h}) and the parameters and definitions in Table~\ref{GEOdata} and \ref{SRparam} we are now able to plot the linear noise spectral density of the ideal GEO\,600 topology for output quadrature fields of arbitrary values of the angle $\zeta$. Fig.~\ref{noise1} shows the two spectral densities $S_h(\zeta=0)$ and $S_h(\zeta=\frac{\pi}{2})$ compared with the SQL (straight solid line). It can be seen that for both quadrature angles the SQL is beaten at frequencies around $30\, {\rm Hz}$. This noise minimum is due to the opto-mechanical resonance (i.e. the optical-spring effect). The second minimum at around $200\,{\rm Hz}$ corresponds to the optical resonance of the SR cavity. This resonance can also beat the SQL when higher reflectivities of the SR mirror $\rho$ are used. For further comparison, the quantum noise limit of a conventional GEO\,600 without signal-recycling is also given (solid line in the upper part of Fig.~\ref{noise1}). The dashed lines represent the two contributions to this (conventional) limit, the uncorrelated white shot noise and the radiation-pressure noise ($\propto f^{-2}$). The limit given here is calculated for a circulating light power of $P=10\,{\rm kW}$ that reaches the SQL at $3\,{\rm Hz}$ and of course can never beat the SQL. It is interesting to note that light powers of around $1\,{\rm MW}$ are needed to shift the conventional limit downwards to get standard quantum noise limited sensitivity at around $100\,{\rm Hz}$ (not shown in Fig.~\ref{noise1}).

In the next two sections we investigate how the sub-SQL spectral noise densities of signal-recycled gravitational wave detectors can be further improved by squeezed light injected into the dark port of the interferometer and by a frequency dependent read-out scheme.

\section{Signal recycling and squeezed light input}
\label{SI}

\begin{figure}[ht!]
  \centerline{\includegraphics[width=8.0cm]{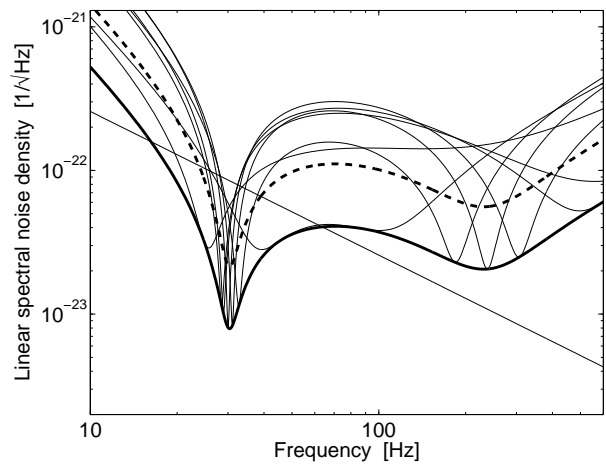}}
  \vspace{0mm}
  \caption{The bold dashed curve shows the phase-quadrature noise spectral density of a SR-interferometer with unsqueezed (coherent) vacuum input. The array of thin black curves evolves from the dashed curve, if the input vacuum-field at the dark port is squeezed with squeezing parameter $r=1$ and the squeezing angle $\lambda$ is varied in a frequency independent manner. The array is bounded from below by the lower bold black curve. Alternatively, one obtains the lower boundary, if the conventional SR noise spectral density is simply shifted downwards by a factor of $\rm{e}^{-r}$. The same holds for the amplitude quadrature. The straight line represents the standard quantum-limit.}
  \label{noise2}
\end{figure}

As first proposed by Caves~\cite{Cav81}, squeezed light can be employed to reduce the high power requirements in GW interferometers. Later Unruh~\cite{Unruh82} and others~\cite{GLe87,JRe90,PCW93,KLMTV01} have found and proven in different ways that squeezed light with a frequency dependent orientation of the squeezing ellipse can reduce the quantum noise down to values beyond the standard quantum limit.
This research was done on interferometer topologies without signal-recycling. Chickarmane {\it et al.}~\cite{CDh96,CDRGBM98} investigated the squeezed-input signal-recycled interferometer at low laser powers, i.e. the shot-noise limited case. In this section we consider the squeezed-input signal-recycled interferometer at high laser powers including the effect of back-action noise.

As discussed in Sec.~IVB of Ref.~\cite{KLMTV01}, squeezed vacuum is related to the ordinary coherent vacuum state by an unitary operator
\beq
|\mbox{in}\rangle = S(r,\lambda)|0\rangle\,,
\eeq
where $r$ is the squeezing parameter, and $\lambda$ the squeezing angle (for an introduction to squeezed light see for example \cite{WallsMilburn}). Alternatively, we can transform the input state back to the vacuum state, by
\beq
|\mbox{in}\rangle \rightarrow S^{\dagger}(r,\lambda)|\mbox{in}\rangle = |0\rangle\,,
\eeq
and at the same time transform the input quadrature operators accordingly [Eq.~(A8) of Ref.~\cite{KLMTV01}],
\bea
\label{sqztransform}
\overline{\mathbf{i}}
&\rightarrow&
S^{\dagger}(r,\lambda)\,
\overline{\mathbf{i}}\,
S(r,\lambda) \nonumber \\
&=&
\mathcal{D}(-\lambda)
\mathcal{S}(r)
\mathcal{D}(\lambda)\overline{\mathbf{i}}\,,
\eea
where
\beq
\mathcal{D}(\lambda)
\equiv 
\left(
\begin{array}{rr}
\cos\lambda & \sin\lambda \\
-\sin\lambda & \cos\lambda
\end{array}
\right)
\,,\quad
\mathcal{S}(r)\equiv 
\left(
\begin{array}{rr}
e^{r} & 0 \\
0 & e^{-r}
\end{array}
\right)\,.
\eeq
From Eq.~(\ref{sqztransform}), we also see that a squeezed vacuum with squeezing angle $\lambda$ can be obtained from a second-quadrature squeezing by applying a rotation of $\mathcal{D}(-\lambda)$ (note the minus sign). Any further rotation of quadratures will also add (with a minus sign) to the squeezing angle. 

The input-output relation of the lossless interferometer with fixed beamsplitter becomes
\beq
\overline{\mathbf{o}}=\frac{1}{M}\left[\mathbf{T} \, 
\mathcal{D}(-\lambda)
\mathcal{S}(r)
\mathcal{D}(\lambda)\overline{\mathbf{i}}
 + \overline{\mathbf{s}}\,h\right]\,,
\eeq
implying a noise spectral density of 
\beq
\label{ShSI}
S_h=
\frac{\left(
\begin{array}{cc}
\cos\zeta & \sin\zeta
\end{array}
\right)
\mathbf{T}\mathcal{D}(-\lambda) \mathcal{S}(2r) \mathcal{D}(\lambda)
\mathbf{T}^{\dagger}
\left(
\begin{array}{c}
\cos\zeta \\ \sin\zeta
\end{array}
\right)}
{\left(
\begin{array}{cc}
\cos\zeta & \sin\zeta
\end{array}
\right)
\overline{\mathbf{s}}\,\overline{\mathbf{s}}^{\dagger}
\left(
\begin{array}{c}
\cos\zeta \\ \sin\zeta
\end{array}
\right)}
\eeq
Note that here $\mathbf{T}$ is a real matrix with an overall phase factor in front (cf.~Eq.~(\ref{TransM})).   
Fig.~\ref{noise2} shows an array consisting of 7 curves (thin lines) where the quadrature angle $\zeta = \pi/2$ is constant and the frequency independent squeezing angle $\lambda$ is varied. In all cases the squeezing parameter $r$ has been set to unity. Interestingly a variation of the frequency-independent squeezing angle causes a frequency shift of both resonances.
For comparison, the standard quantum limit (straight line) and the spectral noise density in the  quadrature at $\zeta = \pi/2$ without squeezed input is also given (dashed line). 
As we can see, each individual frequency-independent value for $\lambda$ can be advantageous to the case without squeezing only in a certain frequency band. 
Obviously, the envelope of the minima of the squeezed input array, as also drawn in the graph (lower bold line), is physically meaningful since it can in principle be realized by applying squeezed light with a squeezing angle optimized for each side-band frequency. Such light is called frequency-dependent squeezed light and yields a broad-band improvement in the quantum noise limited sensitivity. In the final paragraphs of this section we now derive an analytical expression for the optimized spectral noise density.
Suppose now the squeezing angle $\lambda$ can be an arbitrary function of frequency, and $r$ is always positive, then as we can tell from Eq.~(\ref{ShSI}), the optimal $\lambda(\Omega)$ should make 
\beq
\mathcal{D}(\lambda(\Omega)) \mathbf{T}^{\dagger}
\left(
\begin{array}{c}
\cos\zeta \\ \sin\zeta
\end{array}
\right)
\propto 
\left(
\begin{array}{c}
0 \\ 1
\end{array}
\right)\,,
\eeq
or
\beq
\label{desiredlambda}
\tan\lambda(\Omega) =  -\frac{T_{11}\cos\zeta + T_{21}\sin\zeta}{T_{12}\cos\zeta +T_{22}\sin\zeta}\,;
\eeq
yielding an optimal noise spectrum of 
\beq
S_h^{\rm SI}=
e^{-2r}
\frac{\left(
\begin{array}{cc}
\cos\zeta & \sin\zeta
\end{array}
\right)
\mathbf{T}
\mathbf{T}^{\dagger}
\left(
\begin{array}{c}
\cos\zeta \\ \sin\zeta
\end{array}
\right)}
{\left(
\begin{array}{cc}
\cos\zeta & \sin\zeta
\end{array}
\right)
\overline{\mathbf{s}}\,\overline{\mathbf{s}}^{\dagger}
\left(
\begin{array}{c}
\cos\zeta \\ \sin\zeta
\end{array}
\right)}\,.
\label{SIopt}
\eeq
This expression turns out to be identical to the noise spectral density without squeezing in Eq.~(\ref{SRSD}) being suppressed by a factor of $e^{-2r}$. This result can be understood intuitively as follows. The input quadrature field is going to be rotated (and possibly ponderomotively squeezed) by the matrix $\mathbf{T}$ before being detected. The minimal noise quadrature of the squeezed state should therefore be rotated conversely before being injected into the interferometer, such that the detector always ``sees'' the minimal noise. 

Squeezed vacuum can be generated with a variable but frequency-independent squeezing angle $\lambda$ (see for example \cite{BISM98}). 
A frequency-dependent squeezing angle can be obtained subsequently by filtering the initial squeezed light through detuned Fabry-P\'erot (FP) cavities, as proposed by Kimble et al.~\cite{KLMTV01}, which can rotate the quadratures in a frequency dependent way. 
For small frequencies ($\Omega \ll  c/L_{\rm FP}$), a detuned FP cavity of length $L_{\rm FP}$ rotates the reflected quadrature in the following way:
\beq
\overline{\mathbf{a}}^{\rm out}=e^{i\alpha_{\mathrm{m}}}\mathcal{D}(-\alpha_{\mathrm{p}})\overline{\mathbf{a}}^{\rm in}
\eeq
with 
\beq
\alpha_{\mathrm{p,m}}=\frac{1}{2}(\alpha_+ \pm \alpha_-)
\eeq
and
\beq
\alpha_{\pm}=2 \arctan(\xi \pm \Omega/\delta).
\eeq
where $\xi $ is defined by the resonant frequency $\omega_{\rm FP}$ and by $\delta$ which is the half-linewidth of the cavity:  $\omega_{\rm FP}=\omega_0 -\xi\delta$.
As further shown by Purdue and Chen in Appendix A of Ref.~\cite{PCh02}, several such Fabry-P\'erot filter cavities can be combined to give a broad category of frequency dependent rotation angles. 
Adopting their formulas [cf.~Eqs.~(A.8)---(A.14)] into our context, we found that, in order to realize an {\it additional} squeezing angle $\lambda(\Omega)$ with the form of
\beq
\label{possiblelambda}
\tan\lambda(\Omega)=\frac{\sum_{k=0}^{n}B_k \Omega^{2k}}{\sum_{k=0}^{n}A_k \Omega^{2k}}\,,\quad |A_n+i\,B_n|>0\,.
\eeq
we first need to obtain an initial frequency independent squeezed state with  
\beq
\label{initialsqueezing}
\lambda_0 = \arg (A_n -i\,B_n)\,,
\eeq
and then filter this squeezed light with $n$ filters whose complex resonant frequencies differ from $\omega_0$ by $\Omega_J^{\rm res} \equiv - \delta_J \xi_J -i\,\delta_J$, $J=1,2,\ldots, n$, with $\{ \pm \Omega_J^{\rm res} \}$  being the $2n$ roots of  the {\it characteristic equation}:
\beq
\label{FDsqueezing}
\sum_{k=0}^n (A_k - i\,B_k)\Omega^{2k}=0\,.
\eeq
[Note that $\{ \Omega_J^{\rm res} \}$ are the $n$ roots with the appropriate sign of imaginary part, in our case negative.]

Suppose the readout quadrature $\zeta$ is frequency independent, from the ideal input-output relation of GEO\,600, we see that the desired $\lambda$ from Eq.~(\ref{desiredlambda}) is indeed of the form of Eq.~(\ref{possiblelambda}) when $\Omega L/c$ is expanded to the leading order~\footnote{Note that, one has to take $\Omega L/c \sim T_{\rm SR} \sim \phi \sim \mathcal{K}$ in order to get a meaningful expansion.}. Two filter cavities are necessary for the generic case. However, as we look at the low-power limit, only one such filter is necessary. In this case, the input-output relation rotates the input quadratures into the output quadratures following the same law as a detuned cavity. Naturally, as we go through Eqs.~(\ref{initialsqueezing}) and (\ref{FDsqueezing}), we find that the required initial additional squeezing angle is 
\beq
\label{initlambdaGEO}
\lambda_0 = \zeta -\pi/2\,,
\eeq
which puts the minor axis of the noise ellipse onto the $\zeta$ quadrature, while the required cavity has resonant frequency
\beq
\Omega^{\rm filter\,res} = \frac{\phi_{\rm FP} c}{L_{\rm FP}} - i \frac{c \tau_{\rm FP}^2}{4 L_{\rm FP}}\,,
\label{REScompl}
\eeq
which is just ``opposite'' to the signal-recycling resonant frequency,
\beq
\Omega^{\rm SR\,res} \equiv  \omega^{\rm SR} - i\,\gamma^{\rm SR} =- \frac{\phi c}{L} - i \frac{c \tau^2}{4 L}\,,
\eeq
and cancels the rotation induced by signal-recycling. 
\begin{widetext}
For full-power GEO\,600 interferometers, the initial additional squeezing angle is still given by Eq.~(\ref{initlambdaGEO}), while the frequency-dependent part requires two cavities determined by the following characteristic equation: 
\beq
\Omega^2(\Omega+\omega^{\rm SR} +i\,\gamma^{\rm SR})(\Omega-\omega^{\rm SR} -i\,\gamma^{\rm SR})
 - \frac{10 P \omega_0}{m_{\rm M} L c} (\omega^{\rm SR} +2 e^{i\zeta}\sin(\zeta) \gamma^{\rm SR}) =0 \,.
\eeq
\end{widetext}
It is straightforward to solve for the four roots (in two pairs) of the characteristic equations. The corresponding transmissivity $\tau$ of the input mirror and the detuning $\phi$ of the filter cavity can be derived from these roots by virtue of Eq.~(\ref{REScompl}).

\section{Signal recycling, squeezed light input and variational output}
\label{VO}

As shown by Kimble {\it et al.} \cite{KLMTV01}, the  quantum noise spectral density of a conventional interferometer without signal-recycling can benefit simultaneously from both, frequency-dependent squeezed light input and frequency-dependent homodyne read-out. In this section we investigate the optical-spring signal-recycled interferometer with corresponding additional input and output optics. We start from the result of the previous section and vary the angle $\zeta$ of homodyne detection. 

\begin{figure}[ht!]
  \centerline{\includegraphics[width=8.0cm]{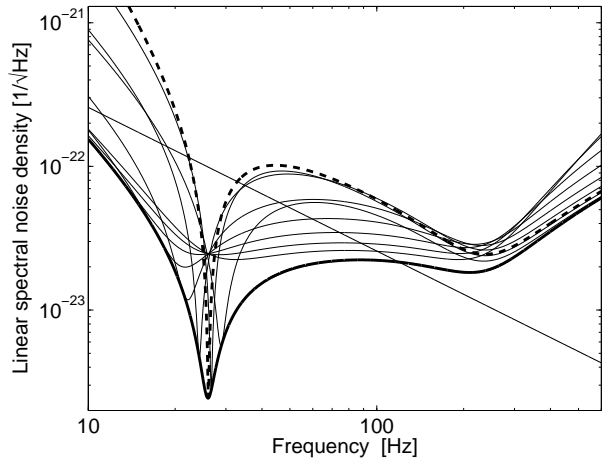}}
  \vspace{0mm}
  \caption{Another improvement of the SR-densities is achieved, if the detection angle is optimized for each signal frequency. Since the shot noise and radiation-pressure noise are highly correlated especially in the detection band, the effect is less beneficial than the optimization of the input squeezing angle. However, comparing the boundary curve with the dashed curve which corresponds to an arbitrary but fixed detection angle, the bandwidth of the noise minima is enhanced and a noise reduction by a factor of 10 can be achieved at some frequencies.}
  \label{noise3}
\end{figure}

Fig.~\ref{noise3} shows an array of noise spectral densities of the output quadrature with varying detection angle $\zeta$ which is here still frequency independent. The input vacuum at the dark port is optimally squeezed with squeezing parameter $r=1$. Obviously, the array is bounded from below. This boundary corresponds to the optimized quantum noise spectral density of the signal-recycled interferometer. 
One member of the array is highlighted by a bold dashed line. Comparing the dashed curve with the optimized noise spectral density, one can see that the variational output provides a further improvement of the interferometer's performance which is mainly an increased bandwidth of the sub-SQL sensitivity. At some frequencies the noise is reduced by a factor of 10. We emphasize that the optimized noise spectrum presented can not be further improved for this interferometer topology without increasing the squeezing parameter $r$ of the input vacuum. Obviously, our results are also significant without any squeezing of the input vacuum. The plots in Fig.~\ref{noise3} are not altered except for a shift upwards by a factor of $e^r$.

In the final part of this section we give an analytical expression of the lower boundary starting from Eq.~(\ref{SIopt}). $S_h^{\rm SI}$ has to be minimized with respect to the detection angle $\zeta$. One method to find the minimum noise is to determine analytically the minimum of the function $S_h^{\rm SI}(\zeta)$. 
Then, a lengthy but straightforward calculation leads to a conditional equation for the optimized detection angle $\zeta_{\rm opt}$ of the following form
\footnote{A similar analytic expression has also been obtained independently by Buonanno and Chen, but remained unpublished.}:
\beq
\left(\begin{array}{cc}\cos\zeta_{\rm opt} & \sin\zeta_{\rm opt}\end{array}\right)
\left(\begin{array}{cc}Q_{\rm 11} & Q_{\rm 12} \\ Q_{\rm 12} & Q_{\rm 22} \end{array}\right)
\left(\begin{array}{c}\cos\zeta_{\rm opt} \\ \sin\zeta_{\rm opt}\end{array}\right)=0.
\label{QUADR}
\eeq
Representing a general SR interferometer, the coefficients of the symmetric quadric $\mathbf{Q}$ are complex (and complex-valued) functions of the interferometers' parameters ($\mathcal{K}, \Phi,\ldots$) which determine the input-output relation Eq.~(\ref{IOrel}). It is more convenient to express them in terms of the elements of the two symmetrized matrices
$\mathcal{S}=\left<\overline{\mathbf{s}}\,\overline{\mathbf{s}}^{\dagger}\right>_{\rm{sym}}$,
$\mathcal{T}=\left<\mathbf{T}\,\mathbf{T}^{\dagger}\right>_{\rm{sym}}$:
\beq
\begin{array}{rcl}
Q_{\rm 11} & = & \mathcal{S}_{11}(\mathcal{T}_{12}+\mathcal{T}_{21})-\mathcal{T}_{11}(\mathcal{S}_{12}+\mathcal{S}_{21}),\\ \\
Q_{\rm 12} & = & \mathcal{S}_{11}\mathcal{T}_{22}-\mathcal{T}_{11}\mathcal{S}_{22},\\ \\
Q_{\rm 22} & = & \mathcal{T}_{22}(\mathcal{S}_{12}+\mathcal{S}_{21})-\mathcal{S}_{22}(\mathcal{T}_{12}+\mathcal{T}_{21}).
\end{array}
\eeq
\begin{figure}[ht!]
  \centerline{\includegraphics[width=8.0cm]{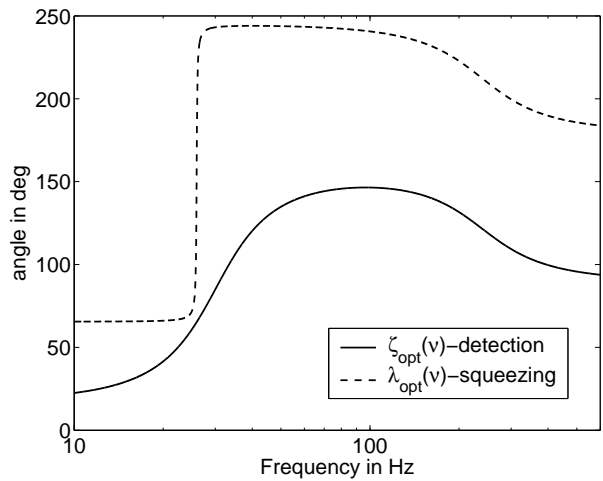}}
  \vspace{0mm}
  \caption{The optimized detection angle $\zeta_{\rm{opt}}$ is determined by Eq.~(\ref{ZETA}). The optimized squeezing angle $\lambda_{\rm{opt}}$ of the input field depends on $\zeta_{\rm{opt}}$ by virtue of Eq.~(\ref{desiredlambda}).}
  \label{OptAngle}
\end{figure}
In general, Eq.~(\ref{QUADR}) has two solutions corresponding to a local minimum and a local maximum of the noise density:
\beq
\zeta_\pm=-\rm{arccot}\left[\frac{1}{Q_{\rm 11}}\left(\pm\sqrt{-\det\mathbf{Q}}+Q_{\rm 12}\right)\right].
\label{ZETA}
\eeq
The minimum of the noise spectral density is given by inserting $\zeta_{\rm opt}=\zeta_-$ into Eq.~(\ref{SIopt}). The optimized detection angle $\zeta_{\rm{opt}}$ is shown in Fig.~\ref{OptAngle} together with the optimized squeezing angle $\lambda_{\rm{opt}}$ of the input field which depends on $\zeta_{\rm{opt}}$ according to Eq.~(\ref{desiredlambda}). The form of both curves suggests that the filtering of the input and output light is accomplishable. Due to the frequency dependence of the squeezing and detection angle, one has to investigate first if an expansion in the form of Eq.~(\ref{possiblelambda}) yields an expression which represents a manageable number of filter cavities. Furthermore, both spectra in Fig.~\ref{OptAngle} are sensitive to small changes of the parameters. Therefore, we do not propose a specific number of filter cavities needed to realize the frequency dependence. 

\section{Conclusion}
\label{CON}
We have shown that the quantum-noise limited sensitivity of signal-recycled interferometers, like GEO\,600 or LIGO\,II, can be improved by additional input and output optics. Although an optical-spring signal-recycled interferometer can already beat the standard quantum limit and ponderomotively generates squeezed light our results show that squeezed light input is compatible leading to a quantum noise reduction by the squeezing factor $e^{-2r}$ (in power). Variational output optics was proven to provide an additional benefit to the quantum noise limited sensitivity of signal-recycled interferometers. Our work augments the results by Buonanno and Chen \cite{BCh01a,BCh01b,BCh02a,BCh03a} and by Kimble {\it et al.} \cite{KLMTV01} and synthesizes their investigations.\\
We have provided fully optimized homodyning and squeezing angles, expressible in analytical formulas, although they are probably not easily achievable using the technique proposed by Kimble {\it et al.}, which uses detuned FP cavities as optical filters. 
However, in the special (sub-optimal) case with frequency-independent homodyning angle but frequency-dependent input squeezing angle, we found the optimal (frequency-dependent) input squeezing angle to be achievable by applying two subsequent filters on a frequency-independent squeezed light. 
We did not analyze the effect of optical losses, but as pointed out by Kimble {\it et al.}, the frequency-dependent input squeezing technique is less susceptible to optical losses than the variational readout and squeezed-variational schemes. A thorough study of optical losses will be published in a separate paper~\cite{Harms03b}.\\

We acknowledge A.~Freise, H.~L\"uck, G.~Heinzel and B.~Willke for many discussions providing us with valuable insight into signal-recycled interferometers. The research of YC is supported by the National Science Foundation grant PHY-0099568 and by the David and Barbara Groce fund at the San Diego Foundation. The author thanks the Albert-Einstein-Institut in Hannover for support during his visit. The author also thanks Alessandra Buonanno for collaboration in numerous earlier works, from which his contributions to this paper benefit.

\appendix 


\begin{thebibliography}{12}
\bibitem{Knote} If one does not make use of the two approximations that the arm cavities are short compared to the wavelength assigned to the modulation frequency $\Omega$ and that the transmissivity $\tau_{\rm arm}$ of the cavity's internal mirror is small, the coupling constant for the Advanced LIGO configuration assumes a form which highlights the distinguishing properties as well as the similarities even more clearly:\\ $\mathcal{K}_{\rm LIGO}=\displaystyle{\frac{32P\omega_0}{mc^2\Omega^2}\frac{1}{\left|1-\sqrt{1-\tau_{\rm arm}^2}{\rm e}^{2\imath\frac{\Omega L}{c}}\right|^2}}$.
\end{thebibliography}
\end{document}